\documentclass[aps,prl,twocolumn,superscriptaddress,floatfix]{revtex4-2}
\usepackage{amsmath}
\usepackage{graphicx}
\usepackage{color}

\newif\ifshowcomments
\showcommentstrue   

\ifshowcomments
  \newcommand{\mr}[1]{\textcolor{red}{#1}}
  \newcommand{\GG}[1]{\textcolor{blue}{#1}}
  \newcommand{\mb}[1]{\textcolor{blue}{#1}}
  \newcommand{\su}[1]{\textcolor{magenta}{#1}}
\else
  \newcommand{\mr}[1]{}
  \newcommand{\GG}[1]{}
  \newcommand{\mb}[1]{}
  \newcommand{\su}[1]{}
\fi

\begin{document}

\title{Bulk and surface electronic structure of MoAlB(010)}
\author{Gianmarco Gatti}
\affiliation{Department of Physics and Astronomy, Aarhus University, 8000 Aarhus C, Denmark}
\author{Amalie H. Svaneborg}
\affiliation{Department of Physics, DTU, 2800, Kgs. Lyngby, Denmark}
\author{Wu Bing}
\affiliation{Department of Inorganic Chemistry, University of Chemistry and Technology Prague, Technicka 5, 166 28 Prague 6, Czech Republic}
\author{Gesa-R. Siemann}
\affiliation{Department of Physics and Astronomy, Aarhus University, 8000 Aarhus C, Denmark}
\author{Anders S. Mortensen}
\affiliation{Department of Physics and Astronomy, Aarhus University, 8000 Aarhus C, Denmark}
\author{Naina Kushwaha}
\affiliation{Central Laser Facility, STFC Rutherford Appleton Laboratory, OX11 0QX, Harwell, UK}
\affiliation{SUPA, School of Physics and Astronomy, University of St Andrews, KY16 9SS, St Andrews, UK}
\author{Jennifer Rigden}
\affiliation {Central Laser Facility, STFC Rutherford Appleton Laboratory, OX11 0QX, Harwell, UK}
\author{Jakob K. Svaneborg}
\affiliation{Department of Physics, DTU, 2800, Kgs. Lyngby, Denmark}
\author{Matthew D. Watson}
\author{Timur K. Kim}
\affiliation{Diamond Light Source Ltd, Harwell Science and Innovation Campus, Didcot, OX110DE, UK}
\author{Charlotte E. Sanders}
\affiliation {Central Laser Facility, STFC Rutherford Appleton Laboratory, OX11 0QX, Harwell, UK}
\affiliation{SUPA, School of Physics and Astronomy, University of St Andrews, KY16 9SS, St Andrews, UK}
\author{Kristian S. Thygesen}
\affiliation{Department of Physics, DTU, 2800, Kgs. Lyngby, Denmark}
\author{Zdenek Sofer}
\affiliation{Department of Inorganic Chemistry, University of Chemistry and Technology Prague, Technicka 5, 166 28 Prague 6, Czech Republic}
\author{Philip Hofmann}
\email{philip@phys.au.dk}
\affiliation{Department of Physics and Astronomy, Aarhus University, 8000 Aarhus C, Denmark}
\date{\today}
\begin{abstract}
The bulk and surface electronic structure of MoAlB(010) is studied by a combination of angle-resolved photoemission spectroscopy and density functional calculations. The observed bulk Fermi-level crossings agree with the previously reported bulk Fermi surface of the material. Additionally, we find several surface states in the wide projected bulk band gaps around the Fermi energy. The surface states differ in their stability under residual-gas exposure in the vacuum system and in the magnitude of their Rashba-type spin--orbit splitting. We explain this in terms of their elemental and orbital character. A surface state arising from Al dangling bonds is sensitive to surface contamination, whereas a mainly Mo-derived surface state exhibits the stronger spin--orbit splitting. The surface states show symmetry-enforced crossings near the $\bar{\mathrm{S}}$ point of the surface Brillouin zone. These are protected by the mirror-symmetry elements of the \textit{p2mm} wallpaper group.
\end{abstract}

\maketitle

\section{Introduction}

MAB phases are ternary compounds composed of transition metals (M), an A-group element (typically from groups 13 or 14, e.g.\ Al or Si), and boron (B) \cite{Kota:2020aa}. They form a family of layered materials closely related to the MAX phases \cite{Dahlqvist:2024aa}, where X denotes carbon or nitrogen. Like MAX phases, MAB compounds combine metallic properties—such as good electrical and thermal conductivity—with ceramic characteristics, including a high Young’s modulus, high melting point, and resistance to oxidation; these make them attractive for a range of applications \cite{Kota:2020aa}.
As in MAX phases, MAB materials crystallize as natural heterostructures in which strongly bonded planar MB units are interleaved with more weakly bonded A layers. This layered bonding hierarchy not only underlies the mechanical and transport properties of MAB phases, but also gives rise to a rich electronic structure. Variations in synthesis conditions allow control over the stacking of the structural units. The chemical difference between the A and MB layers can be exploited to transform MAB crystals into two-dimensional MBenes via etching \cite{Ramezanzadeh:2025aa}.

The crystal structure of the MAB phase MoAlB is shown in Fig.~\ref{fig:1}(a) and (b). MoAlB crystallizes in the non-symmorphic \textit{Cmcm} space group \cite{Kota:2016aa}. The crystal structure is conventionally described using the orthorhombic unit cell indicated by the black lines in the figure. Alternatively, a unit cell of half the size can be chosen (red lines). We refer to these unit cells as the conventional cell and the primitive cell, respectively, and we use the conventional cell unless stated otherwise. The MoAlB structure consists of layers of side-sharing BMo$_6$ trigonal prisms that alternate with double Al square lattices along the crystallographic $b$ axis. In contrast to many MAX carbides and nitrides, MoAlB exhibits a pronounced in-plane structural anisotropy arising from zig-zag B chains running along the $a$ axis. Together with the out-of-plane anisotropy imposed by the layered structure, this leads to a strong three-dimensional anisotropy in the electronic, vibrational, and transport properties \cite{Huang:2021aa}.
Density functional theory calculations indicate a predominantly covalent character of the Mo--B and B--B bonds, whereas the Mo--Al and Al--Al bonds are largely metallic in nature. The electronic states forming the Fermi surface are mainly derived from Mo and Al orbitals, and the structural anisotropy is reflected in a highly anisotropic, largely two-dimensional Fermi surface comprising many sheets according to both calculations \cite{Li:2016ac,Ali:2017aa,Ke:2017aa,Huang:2021aa} and quantum oscillations \cite{Zhao:2020ad,Nie:2022aa}.

\begin{figure}[t]
 \includegraphics[width=0.45\textwidth]{Fig1.pdf}\\
\caption{(Color online) Crystal structure of MoAlB. 
(a) Side view of the $ab$ plane. The orthorhombic conventional unit cell is indicated in black, while the primitive cell is shown in red. The natural cleavage plane is indicated by the horizontal blue line.
(b) Three-dimensional view of the two principal building blocks of the layered unit cell, namely the Al square lattice and the BMo$_6$ covalent layer.
(c) Sketch of the MoAlB Brillouin zone (BZ). The conventional, primitive, and (010) surface-projected BZs are marked in black, red, and orange, respectively.
}
  \label{fig:1}
\end{figure}

An additional important aspect of MoAlB is its non-symmorphic \textit{Cmcm} space group symmetry, which has direct consequences for both the bulk and surface electronic structure. 
The non-symmorphic character of MoAlB derives from the presence of glide planes and screw axes defined when considering the primitive unit cell.
The common choice to adopt the conventional unit cell introduces additional symmetries, which are combinations of a point-group symmetry and fractional translations along the $b$ axis. We stress that these symmetry operations are, however, not proper non-symmorphic operations, since the involved translations can be described by the translation vectors spanning the primitive lattice. Overall, this leads to a situation where the use of a non-primitive unit cell not only results in an apparently increased number of electronic states but also introduces apparent non-symmorphic symmetries. 

A sketch of the bulk Brillouin zones (BZs) for both the conventional cell (black solid lines) and the primitive cell (red solid lines) is given in Fig.~\ref{fig:1}(c). The panel also shows the projection of the conventional and primitive BZs onto the (010) surface, resulting in an identical surface BZ (orange).

In this paper, we present a comprehensive investigation of the bulk and surface electronic structure of MoAlB(010) using a combination of angle-resolved photoemission spectroscopy (ARPES) and first-principles calculations. The measurements confirm the strongly metallic character of MoAlB and are in agreement with previous calculations and experimental studies of its Fermi surface. A central result of this work is the identification of strongly surface-localized electronic states within wide projected bulk band gaps and the characterization of these states with respect to their elemental and orbital character, as well as their symmetry.

\section{Methods}

Millimeter-sized MoAlB single crystals were synthesized using a flux method via the following procedure. Mo powder (300 mesh), Al powder (300 mesh), and amorphous B powder were used as starting materials. The raw materials were mixed according to a molar ratio of Mo:Al:B = 3:21:1. The excess Al and Mo were introduced to act as an Al and Mo--Al alloy flux. The mixed powders were loaded and heated in a furnace under an Ar atmosphere. The temperature was first increased to 1550~$^\circ$C at a rate of 5~$^\circ$C/min and maintained for 4 hours, followed by cooling to 1200~$^\circ$C at 1~$^\circ$C/min, and then further cooling to room temperature at 10~$^\circ$C/min.
After synthesis, the obtained solidified product was placed in 3~mol/L HCl and heated to 60~$^\circ$C to dissolve most of the flux. The remaining residual flux was then rapidly removed using 9~mol/L HCl. Finally, the resulting MoAlB crystals were alternately washed with deionized water and acetone, and then dried. The resulting crystals are brittle and typically exhibit large, flat facets with a metallic luster.

Micro-ARPES experiments were performed at the I05 beamline of Diamond Light Source. Samples were mounted using silver epoxy, top-posted, and cleaved \textit{in situ} at a temperature of approximately 25~K. ARPES data were acquired at 25~K using a focused light spot with a diameter of approximately 5~$\mu$m. The light polarization was linear horizontal. The angular and energy resolutions were 0.2$^\circ$ and 30~meV, respectively. The pressure during the measurements was 1.0$\cdot$10$^{-10}$~mbar.
 
Electronic structure calculations were performed using density functional theory (DFT) as implemented in the GPAW code \cite{Mortensen:2024aa}. The exchange--correlation potential was treated within the Perdew--Burke--Ernzerhof approximation \cite{pbe}. Both bulk and slab geometries were initially constructed using the Atomic Simulation Environment \cite{ase-paper}, employing lattice parameters taken from Ref.~\cite{Kota:2016aa}. The slab model consists of two out-of-plane layers of the conventional bulk unit cell. Both bulk and slab structures were fully relaxed and subsequently symmetrized according to the $Cmcm$ and $P2_1/m$ space groups, respectively.

\section{Results and Discussion}

\begin{figure*}[t]
 \includegraphics[width=1 \textwidth]{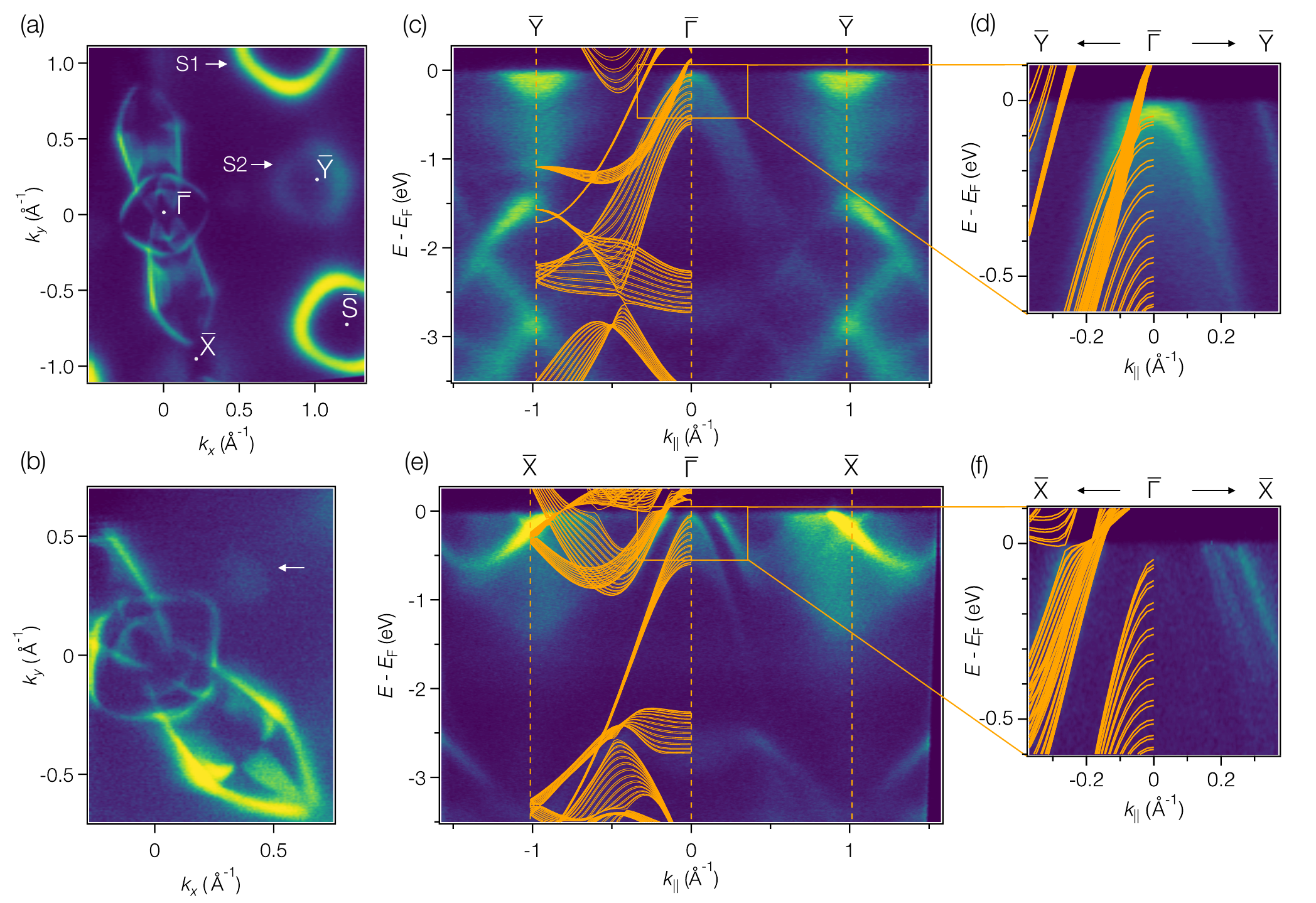}\\
\caption{(Color online) Electronic structure of MoAlB(010) determined by ARPES. 
(a) Photoemission intensity at the Fermi energy measured with a photon energy of $h\nu=$~62~eV. S1 and S2 denote the Fermi contours generated by surface states. 
(b) Photoemission intensity at the Fermi energy measured with $h\nu = 32.5$~eV in selected regions of the surface Brillouin zone. The arrow highlights a weak feature that is assigned to a three-dimensional Fermi surface element. 
(c) Band dispersion along the $\bar{\mathrm{Y}}$--$\bar{\Gamma}$--$\bar{\mathrm{Y}}$ direction measured with $h\nu = 100$~eV and corresponding projected DFT bulk band structure. Each line shown in the calculated dispersion corresponds to a different value of $k_{\perp}$.
(d) Close-up of panel (c) in an energy--momentum region near the Fermi energy and the $\bar{\Gamma}$ point, measured with $h\nu = 32.5$~eV.
(e) Same as panel (c), but along $\bar{\mathrm{X}}$--$\bar{\Gamma}$--$\bar{\mathrm{X}}$.
(f) Close-up of panel (e), but measured with $h\nu = 48$~eV.
The data shown in panels (c) and (d) have been symmetrized around $\bar{\Gamma}$ for illustrative purposes.
}
  \label{fig:2}
\end{figure*}

\subsection{Surface termination}

The weakest bonding in the MoAlB structure occurs between adjacent Al layers, suggesting that the natural cleavage plane lies between these layers, perpendicular to the crystallographic $b$ axis. Cleaving the crystal to form the MoAlB(010) surface is thus expected to result in a single surface termination, as opposed to, \textit{e.g.}, a mixture of Al and Mo terminations.
To verify this, we examine the spatial homogeneity of the Al~2$p$ and Mo~4$p$ core level spectra, as well as the valence-band signal, across the sample surface. Over lateral distances of several hundred~$\mu$m, the spectra show only minor variations, indicating a largely homogeneous surface. Throughout the scanned area, the Al~2$p$ core level signal is systematically much more intense than the Mo~4$p$ signal, with an intensity ratio of $\approx 500$ (for the core level spectra see the online supplementary material (SM) \cite{supp}).
At the employed photon energy of 100~eV, the ratio of the corresponding photoionization cross sections is $\approx 15$, and therefore cannot account for the observed intensity difference. The pronounced dominance of the Al~2$p$ emission strongly indicates that the surface-sensitive photoemission signal is dominated by Al atoms. Small residual spatial variations are attributed to local structural imperfections introduced during the cleavage process. Overall, these observations are consistent with a single, Al-terminated MoAlB(010) surface.

\subsection{Bulk electronic structure}

The surface BZ of MoAlB(010), sketched in Fig.~\ref{fig:1}(c), is a rectangle with nearly equal side lengths; its dimensions are $\bar{\Gamma}-\bar{\mathrm{X}} = 1.013$~\AA$^{-1}$ and $\bar{\Gamma}-\bar{\mathrm{Y}} = 0.978$~\AA$^{-1}$. First-principles calculations predict a bulk Fermi surface composed of six sheets with predominantly two-dimensional character, extending mainly along the $\Gamma$--X direction of the conventional bulk Brillouin zone \cite{Ali:2017aa,Huang:2021aa}.
Figure~\ref{fig:2}(a) displays the photoemission intensity at the Fermi energy $E_\mathrm{F}$ throughout the surface BZ, and Fig.~\ref{fig:2}(b) shows a zoom-in of the bulk Fermi surface structures measured with a lower photon energy. The data confirm the presence of the predicted manifold of bulk Fermi-level crossings near $\bar{\Gamma}$ and along the $\bar{\Gamma}$--$\bar{\mathrm{X}}$ direction. In contrast, the circular features observed around the $\bar{\mathrm{S}}$ and $\bar{\mathrm{Y}}$ points do not follow the predicted bulk dispersion \cite{Ali:2017aa,Huang:2021aa}. The simplest explanation for this discrepancy is to attribute these features to electronic surface states formed after sample cleavage. This scenario is discussed in the following subsection.

Here we focus on the bulk electronic structure. Figures~\ref{fig:2}(c) and (e) present an overview of the electronic dispersion along the $\bar{\Gamma}$--$\bar{\mathrm{Y}}$ and $\bar{\Gamma}$--$\bar{\mathrm{X}}$ directions, respectively, over a broad energy range. Zoomed-in details around $\bar{\Gamma}$ measured with lower photon energies are shown in panels (d) and (f). Several well-defined quasiparticle bands can be seen, including various hole-like bands crossing $E_\mathrm{F}$ along $\bar{\Gamma}$--$\bar{\mathrm{X}}$ (see panel (f)), and an apparent Dirac-like band crossing at the $\bar{\mathrm{X}}$ point near $E_\mathrm{F}$ (panel (e)). Other features are substantially broader, making it difficult to unambiguously trace the dispersion of the visible bands and to identify the potentially numerous Fermi-level crossings. One example is a wide region of high photoemission intensity at $E_\mathrm{F}$ along $\bar{\Gamma}$--$\bar{\mathrm{X}}$ close to $\bar{\mathrm{X}}$ (panel (e)). A similar strong broadening can be observed for the completely filled hole-like band at $\bar{\Gamma}$ just below $E_\mathrm{F}$.

Comparing the data to DFT calculations does in principle require knowledge of the crystal momentum $k_{\perp}$ perpendicular to the surface, which is not conserved in the photoemission process \cite{Plummer:1982aa} and, moreover, suffers from a significant uncertainty due to the small inelastic mean free path of the electrons \cite{Hofmann:2013ab}. In the present case, this is particularly severe because the small size of the bulk BZ in the $k_{\perp}$ direction ($\approx 0.44$~\AA$^{-1}$) means that even a small $k_{\perp}$ broadening can correspond to an integration over the entire BZ in that direction. In order to account for this, we compare the data not to a calculation for a particular $k_{\perp}$ but to the projection of the bulk band structure onto the (010) surface, using only a small number of $k_{\perp}$ points in order not to obscure the experimental data. Such a calculation is overlaid on the left-hand side of the experimental spectra shown in Fig.~\ref{fig:2}(c) to (f). Full bulk band structure calculations for the primitive and conventional unit cells are given in the SM \cite{supp}.

The agreement between ARPES results and calculations near $E_\mathrm{F}$ is very good. The Dirac-like crossing near $\bar{\mathrm{X}}$ is well described by DFT. The observed hole-like structure just below $E_\mathrm{F}$ at $\bar{\Gamma}$ also matches the continuum of calculated bands extremely well, indicating $k_{\perp}$ smearing across the entire BZ for this band. This $k_{\perp}$ smearing also explains the high photoemission intensity at $E_\mathrm{F}$ near $\bar{\mathrm{X}}$, which is due to multiple Fermi-level crossings at different $k_{\perp}$ values. Some of the calculated bands, on the other hand, show very little $k_{\perp}$ dispersion. These appear as sharp bands, in contrast to the broad continua of states elsewhere. Note, for example, the sharp hole-like feature at $\bar{\Gamma}$ that crosses the Fermi level at a wave vector $k \sim 0.2$~\AA$^{-1}$ along the $\bar{\Gamma}$--$\bar{\mathrm{X}}$ direction. Other examples can be found near the $\bar{\mathrm{Y}}$ point: The two bands crossing at $E - E_\mathrm{F} \approx -1.5$~eV, for instance, are essentially two-dimensional. The same is true for the band at $-1$~eV, but this loses its two-dimensional character as it approaches $\bar{\Gamma}$. Consistent with this, in the experiment it is only seen as an identifiable feature near $\bar{\mathrm{Y}}$.

The dense coverage of an area in $E$--$k$ space by calculated bands is not necessarily due to $k_{\perp}$ dispersion. It can also be caused by the presence of many nearly two-dimensional bands in that region. This is almost certainly the case for the Fermi-level crossings close to $\bar{\Gamma}$ in the $\bar{\Gamma}$--$\bar{\mathrm{X}}$ direction in Fig.~\ref{fig:2}(f). While the calculated bands appear to form a broad continuum, the experimental results on the right-hand side show at least four sharp bands, giving rise to well-defined quasi-two-dimensional Fermi surfaces, which are also seen in Fig.~\ref{fig:2}(b). A Fermi surface consisting of many sheets of predominantly two-dimensional character, extended along the $\Gamma$--X direction of the bulk Brillouin zone, is also in very good agreement with earlier DFT calculations \cite{Ali:2017aa,Huang:2021aa}. Indeed, these calculations predict all Fermi surface elements except one to be nearly ``tubes'' in the $\Gamma$--Z direction of the conventional zone. The only Fermi surface element with a distinct three-dimensional character is a pocket along the $\bar{\Gamma}$--$\bar{\mathrm{Y}}$ direction (sheet 4 in Ref.~\cite{Ali:2017aa} and band 5 in Ref.~\cite{Huang:2021aa}). This pocket is also observed here in ARPES, not as distinct crossings but as a weak feature marked with a white arrow in Fig.~\ref{fig:2}(b). Its three-dimensional character is indicated by the fact that this pocket is visible when measured with 32.5~eV photon energy but is not detectable with 62~eV light. The diffuse ARPES signal of this feature can be ascribed to $k_{\perp}$ smearing.

\subsection{Surface electronic structure}

\begin{figure*}[t]
 \includegraphics[width=1 \textwidth]{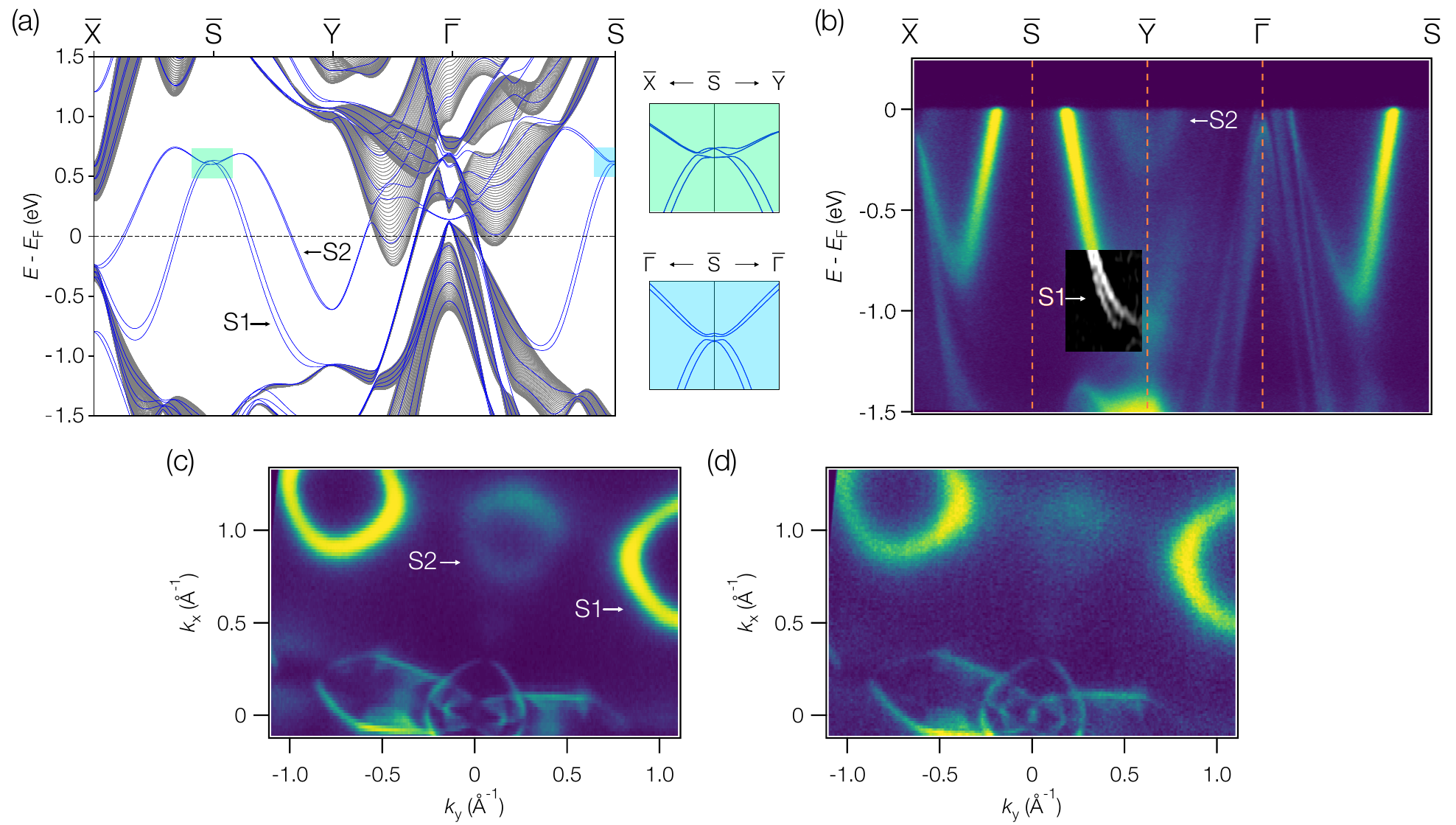}\\
\caption{(Color online) Surface electronic structure of MoAlB(010). 
(a) Projection of the bulk band structure onto the (010) surface (grey), shown together with calculated surface state dispersion including spin--orbit coupling (blue). Two insets on the right-hand side show the surface state dispersion near the $\bar{\mathrm{S}}$ point.
(b) Corresponding measurement along high-symmetry lines of the surface Brillouin zone taken with $h\nu=$~62~eV. The inset shows the curvature of the photoemission intensity to reveal a small splitting of the band.
(c-d) Photoemission intensity at the Fermi energy measured 4 and 8.5 hours after cleaving the sample, respectively ($h\nu=$~62~eV).
The two spectra are normalized to the maximum intensity observed in the whole momentum window.
}
  \label{fig:3}
\end{figure*}

A closer comparison of our ARPES data with bulk band structure calculations in Fig.~\ref{fig:2}(c) reveals that not all experimental features are accounted for by the bulk calculations. At $\bar{\mathrm{Y}}$, a shallow electron pocket near $E_\mathrm{F}$, corresponding to the feature giving rise to the Fermi contour around the $\bar{\mathrm{Y}}$ point in Fig.~\ref{fig:2}(a), and a cross-like structure about 3~eV below $E_\mathrm{F}$ clearly fall outside the bulk band projection. This provides strong experimental evidence for the surface-localized character of these states. This assignment is also supported by DFT.
Figure~\ref{fig:3}(a) shows the projection of the bulk band structure onto the (010) surface (grey), together with the calculated surface state dispersion (blue). Both calculations include spin--orbit coupling. A similar plot over a larger energy range is available in Fig.~S5 of the SM \cite{supp}. The calculated surface bands closely match the experimentally observed dispersions shown in Fig.~\ref{fig:3}(b), confirming that the previously discussed features are indeed surface states.

\begin{figure*}[t]
 \includegraphics[width=1 \textwidth]{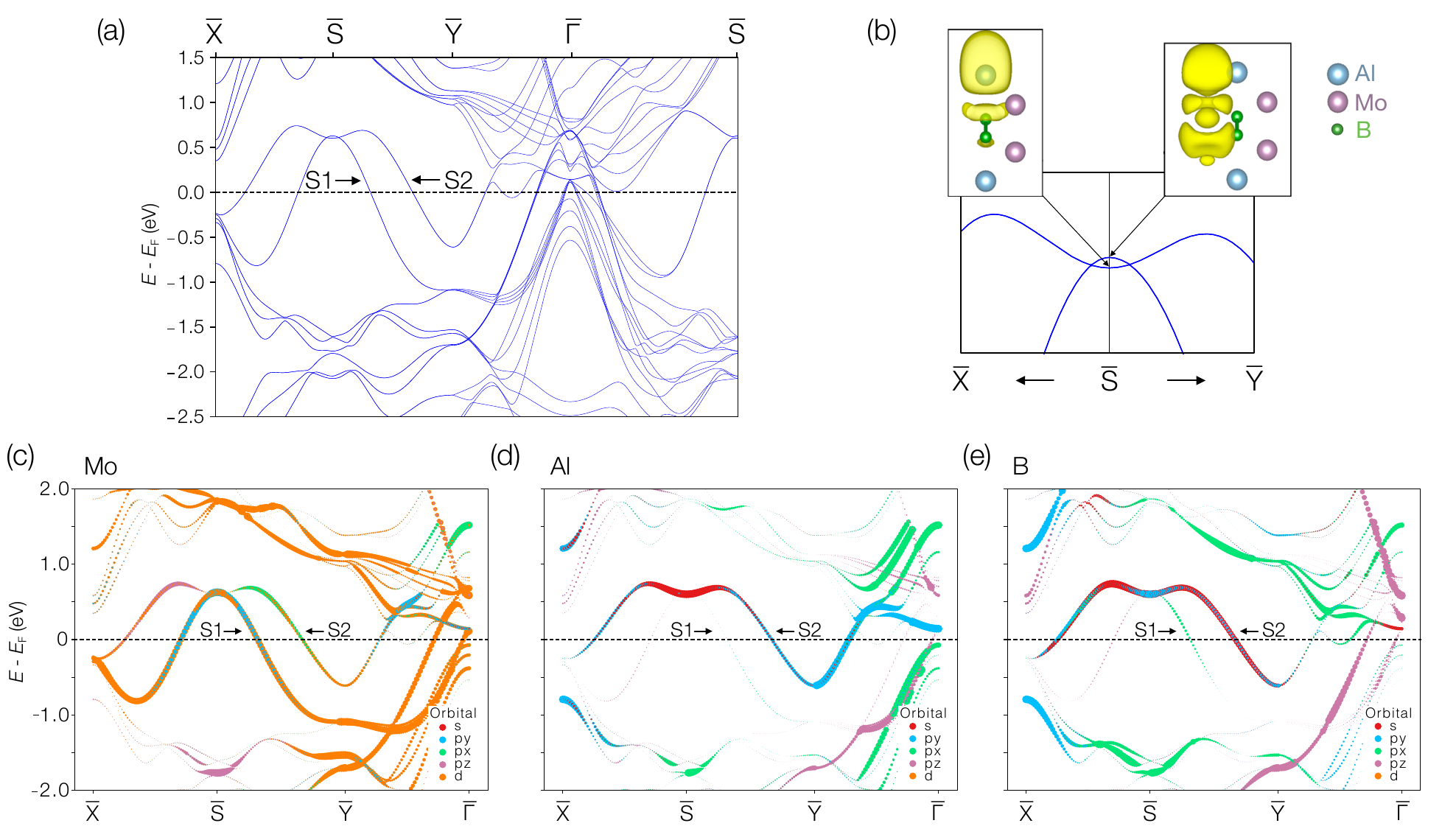}\\
\caption{(Color online) 
(a) Band structure of a four-layer slab of MoAlB calculated without spin--orbit coupling, showing the surface state dispersion around $\bar{\mathrm{S}}$.
(b) Close-up of the dispersion near the band maximum at $\bar{\mathrm{S}}$, together with the corresponding wave function probability density for the S1 and S2 states.
(c)--(e) Orbital-resolved weights of the surface states for near-surface Al, B, and Mo atoms, respectively.
}
  \label{fig:4}
\end{figure*}

We now focus on the surface states crossing $E_\mathrm{F}$, forming the Fermi contours around the $\bar{\mathrm{S}}$ and $\bar{\mathrm{Y}}$ points in Fig.~\ref{fig:2}(a). The calculations in Fig.~\ref{fig:3}(a) reveal that they consist of two pairs of nearly degenerate bands. The first pair connects to the occupied bulk states near the $\bar{\mathrm{X}}$ and $\bar{\mathrm{Y}}$ points and gives rise to a hole pocket centered at $\bar{\mathrm{S}}$. The second pair emerges from the occupied bulk states at $\bar{\mathrm{X}}$ and disperses toward the unoccupied bulk manifold near $\bar{\Gamma}$, forming an electron pocket around the $\bar{\mathrm{Y}}$ point. We refer to these two surface state pairs as S1 and S2, respectively.
The energy splitting between the two bands forming each pair is larger for the S1 state than for the S2 state. For S1, the splitting becomes experimentally resolvable at higher binding energies near the $\bar{\mathrm{Y}}$ point, where it reaches values of up to $\approx 100$~meV. This is illustrated in the inset of Fig.~\ref{fig:3}(b), which shows the curvature of the photoemission intensity to enhance the band contrast \cite{Zhang:2011aa}. 
The energy splitting of the S2 state is not resolved in the data.
The splitting observed in the surface states is attributed to a Rashba-type effect, arising from the combination of spin--orbit coupling and inversion-symmetry breaking at the surface \cite{LaShell:1996aa,Koroteev:2004aa}. This interpretation is supported by calculations performed without spin--orbit coupling, in which the splitting is absent, as shown in Fig.~\ref{fig:4}(a).

A pronounced difference between the S1 and S2 surface states is their different sensitivities to surface contamination by the residual gas in the vacuum system. While the S1 state remains stable over extended periods, exhibiting only a slight increase in spectral linewidth, the S2 state is observable only for a short time after cleaving the sample. This time-dependent suppression of the S2 signal is illustrated by the two constant energy maps at $E_\mathrm{F}$ shown in Fig.~\ref{fig:3}(c)--(d). These measurements were performed under identical experimental conditions, 4.5 hours apart.

The different nature of the S1 and S2 surface states is clarified by examining their atomic and orbital character in the DFT calculations. We consider the surface state dispersion calculated without spin--orbit coupling, shown in Fig.~\ref{fig:4}(a). A magnified view around the $\bar{\mathrm{S}}$ point is given in Fig.~\ref{fig:4}(b), together with the surface state wave function probability density at $\bar{\mathrm{S}}$. From the probability density, we see that the S2 state exhibits a wave function localized at the Al atomic layer of the surface, with some contribution from the underlying B atom, lending it primarily an Al dangling-bond-like character. In contrast, the S1 state is mainly derived from Mo $4d$ orbitals and shows a more delocalized character. This picture is confirmed by plotting the atomic-orbital-resolved spectral weights $w_{nk}^a$ of the surface states in Fig.~\ref{fig:4}(c)--(e). These are computed by projecting the Bloch states $\psi_{nk}$ onto atomic orbitals $\phi_a$, yielding weights $w_{nk}^a = |\langle \phi_a | \psi_{nk} \rangle|^2$. The weights are normalized to the maximum value across all orbitals within each band, such that relative orbital contributions within a given band can be compared, while absolute intensities between different bands cannot. The weights confirm what can be inferred from the wave function plots: S1 is mainly derived from Mo states, while S2 has strong Al character with a B admixture.

From this analysis, a clear distinction between the two surface states emerges, providing a plausible explanation for their markedly different chemical stability. Simple metal surfaces such as Al are well known to be highly reactive, even under ultra--high vacuum conditions, where residual gases---most notably water---can lead to chemical modification of under-coordinated surface sites \cite{Paul:1986aa,Yang:2013ac}. The pronounced dangling-bond character of the S2 state is therefore consistent with its rapid degradation after cleavage. Al, being the lighter element, S2 is also expected to show the  smaller spin--orbit splitting. By contrast, the higher stability of the Mo-derived S1 state suggests that, at the low temperatures of the experiment, surface degradation does not proceed appreciably beyond the Al surface termination. At elevated temperatures, it has been suggested that MoAlB may be protected against oxidation by the segregation of Al atoms to the surface and the formation of a stable, passivating aluminum oxide layer, analogous to the behavior of pure aluminum \cite{Kota:2016aa}.

Despite the reduction of symmetry at the surface, certain crystalline symmetries remain and continue to constrain the surface state dispersion. In particular, the (100) and (001) mirror symmetries of the \textit{p2mm} wallpaper group of the MoAlB(010) surface protect the crossings of the surface states along the $\bar{\mathrm{S}}$--$\bar{\mathrm{X}}$ and $\bar{\mathrm{S}}$--$\bar{\mathrm{Y}}$ high-symmetry lines, as shown in the magnified dispersions in Fig.~\ref{fig:3}(a). For wave vectors along these directions, the little group is isomorphic to the $C_s$ point group. Surface states S1 and S2 belong to the $A'$ and $A''$ irreducible representations of this group, respectively; thus, the Hamiltonian matrix elements between them vanish, and a band crossing is allowed.
Along the $\bar{\mathrm{S}}$--$\bar{\Gamma}$ direction, by contrast, only the identity operation remains in the little group, so that both states belong to the same trivial representation and a small gap opens due to hybridization between the surface-state branches. Notably, the avoided-crossing gap remains very small, which we attribute to the markedly different elemental and orbital character of the involved states, as discussed above. The dispersion in Fig.~\ref{fig:4} shows that this general behavior is also found when spin--orbit coupling is not included: Again, the surface-state dispersions cross along $\bar{\mathrm{S}}$--$\bar{\mathrm{X}}$ and $\bar{\mathrm{S}}$--$\bar{\mathrm{Y}}$, but form a small gap along $\bar{\mathrm{S}}$--$\bar{\Gamma}$.

\section{Conclusion}

We have presented a combined experimental and theoretical investigation of the bulk and surface electronic structure of MoAlB(010). The bulk exhibits a clear metallic character and Fermi surfaces in agreement with our calculations, previous theoretical work, and experimental results from quantum oscillations. The (010) surface supports metallic surface states within wide projected bulk band gaps. At the Fermi energy, two such surface states, S1 and S2, are observed around the $\bar{\mathrm{S}}$ and $\bar{\mathrm{Y}}$ points. A detailed analysis of their orbital character reveals that S1 and S2 are mainly derived from Mo and Al atoms near the surface, respectively. This is consistent with their different sensitivity to residual-gas exposure in the vacuum chamber and with the magnitude of their Rashba spin--orbit splitting. The surface-state degeneracy along certain high-symmetry directions is protected by the residual mirror symmetries of the (010) surface lattice.

\begin{acknowledgments}
This work was supported by Independent Research Fund Denmark  (Grant No. 4258-00002B) and Novo Nordisk Foundation (Grant number NFF23OC0085585). We gratefully acknowledge travel support from DANSCATT and Diamond Light Source. K.S.T. is a Villum Investigator supported by VILLUM FONDEN (grant no. 37789). Z.S. was supported by ERC-CZ program (project LL2101) from Ministry of Education Youth and Sports (MEYS). Z.S. The authors acknowledge the assistance provided by the Advanced Multiscale Materials for Key Enabling Technologies project, supported by the Ministry of Education, Youth, and Sports of the Czech Republic. Project No. CZ.02.01.01/00/22\textunderscore008/0004558, Co-funded by the European Union. We acknowledge Diamond Light Source for time on Beamline I05-1 under Proposals SI39998. 
\end{acknowledgments}

\bibliographystyle{apsrev}

\end{document}